The Meissner and Mesoscopic Superconducting States in 1-4 Unit-Cell FeSe-Films up to 80 K

L. Z. Deng[1], B. Lv[1], Z. Wu[1], Y. Y. Xue[1], W. H. Zhang[2], F. H. Li[2], L. L. Wang[3], X. C. Ma[3], Q. K. Xue[2] and C. W. Chu[1,4]

[1] Texas Center for Superconductivity and Department of Physics, University of Houston, Texas

[2] Department of Physics, Tsinghua University, Beijing

[3] Institute of Physics, Chinese Academy of Sciences, Beijing

[4] Lawrence Berkeley National Laboratory, Berkeley, California

Recent reports [1,2,3] of interface-induced superconductivity in the unit-cell films of FeSe on $SrTiO_3$ with a $T_c$ up to ~ 65±5 K, highest among the Fe-based superconductors and second only to that of the cuprates, have generated great excitement. Here we show results of the first magnetic and resistive investigation on the 1-4 unit-cell FeSe-films on $SrTiO_3$. The samples display the Meissner state below ~ 20 K with a penetration field as low as 0.1 Oe due to large edge effect at 2 K, a mesoscopic superconducting state comprising patches of dimension < 1 μm up to 45~55 K and an unusual relaxation of the diamagnetic signal up to 80 K, suggesting possible superconductivity at this high temperature. The observations demonstrate the heterogeneous nature of the superconducting state in these films and offer new insights into the role of interfaces in high temperature superconductivity proposed and future superconducting electronics fabricated.

The recent reports of unusually high $T_c$ in single-unit-cell (1UC) FeSe films on STO-substrate offer unique opportunities to address issues of high temperature superconductivity, and interfacial superconductivity in particular. Interface-mechanism through phonon-softening at the interface or exchange of excitons across the interface to enhance $T_c$ has long been proposed and explored [4-7]. Interface-enhanced-$T_c$s have been reported [8] in past decades in various artificially made heterostructural layered samples of different materials that are either not superconducting or superconducting with lower $T_c$s by themselves in their equilibrium stoichiometric state at ambient. Unfortunately, questions remain as to the effects of local stress and local doping [9]. Indirect evidence has also been suggested for the unusually high $T_c$ in naturally occurring single crystals [10-12]. Recently, FeSe-films of single unit-cell thick only grown artificially on a STO substrate by the MBE technique were reported [1,2,3] to display a $T_c$ up to 65 K in comparison with the $T_c$ of 8 K of bulk FeSe, representing the clearest demonstration of the interfacial effect on $T_c$-enhancement to date. However, until now, the evidence for such a high $T_c$ comes from the appearance of an energy gap from STM/ARPES measurements, and a resistance drop from a preliminary study. It has been shown [13] that a non-superconductive energy gap can also exist in a strongly correlated electron system such as FeSe. In spite of the extensive global effort, $T_c$ up to 60±5 K in 1UC FeSe films has been reported by only two groups at Tsinghua and Fudan Universities, respectively, showing the delicate nature of these high $T_c$ films and the challenges in their fabrication. We have therefore carried out a systematic magnetic and resistive investigation to look for the Meissner effect, the sufficient condition for superconductivity, the highest possible $T_c$, and also mesoscopic structures in these delicate ultra-thin FeSe films.

We have measured the dc and ac magnetic susceptibilities of eight MBE-grown FeSe films of thickness 1 – 4 unit cells (UC) and the four-lead resistance of all FeSe films from Xue's group in Tsinghua University in Beijing in temperatures up to 300 K and fields up to 7 T, using the Quantum Design MPMS and PPMS systems. These samples consist of nUC FeSe-layers, with n = 1, 3, or 4, on a STO-substrate, protected by layers of Si/10UC FeTe, i.e. Si/10UC FeTe/nUC FeSe/STO of film dimensions 2 mm x 5 mm. Because of the small mass of the FeSe-films which is estimated to be n x 26 ng, special techniques have been developed to acquire the intrinsic results of the ultra-thin FeSe films reproducibly (S1). We have determined the magnetic signals of the protective layer Si/10UC FeTe and the STO-substrate used for the samples. Their values are too small to influence the results of the FeSe films that can be subtracted from the measured value of the samples. All samples exhibit similar magnetic behavior qualitatively but not quantitatively. We shall display the representative magnetic data mainly of the 3 and 4UC-samples due to their stronger superconducting signals at 2 K and their high onset temperatures of diamagnetic shift at ~ 45 and ~ 80 K, respectively. The dc magnetic susceptibilities $\chi_{dc}$s ≡ $M_{dc}$/H during the zero-field-cooled (ZFC) and field-cooled (FC) modes are shown in Fig. 1. A Meissner state below ~ 20 K is clearly evident from the diamagnetic FC-$\chi_{dc}$-shift, resulting in the expulsion of magnetic field from the sample. The ZFC-$\chi_{dc}$ displays a larger diamagnetic shift below 20 K as expected. The relatively large difference between the ZFC-$\chi_{dc}$ and FC-$\chi_{dc}$ suggests a large critical field $H_{c2}$ of the ultra-thin FeSe film. To reduce the scattering of the data, we measured the ac susceptibility $\chi'_{ac}$ ≡ M'/H, where M' is the real part of magnetization ($M_{ac} = M'_{ac} + iM''_{ac}$) at 1.5 kHz, whose magnitude is larger than that of the ZFC-$\chi_{dc}$ and reveals a clear although small diamagnetic shoulder below ~ 45 K followed by a fast drop below 20 K as shown in Fig. 2A. In Fig. 2A, magnetic field is also shown to push both the diamagnetic shoulder and the 20 K-transition to lower temperatures, and eliminate completely the shoulder above 0.1 T, demonstrating that superconductivity exists below 45 K in the 3UC sample. The 4UC-sample shows similar behavior but with a diamagnetic shoulder starting at ~ 80 K, although a smaller dc superconducting signal at low temperature as shown in Fig. 2B. To seek possible other signatures of superconductivity at such a high temperature, we measured the four-lead resistance R at different T's of the 1, 3, and 4UC samples in different fields. As exemplified in Fig. 3 for the 4UC sample, dR/dT shows two peaks signaling two superconducting transitions centering around ~ 18 K and ~ 28 K, respectively. Both are suppressed by the magnetic field. The deviation of dR/dT from a constant below ~ 80 K suggests the possible existence of superconductivity up to ~ 80 K in the 4UC sample, in agreement with the $\chi'_{ac}$(T) shown in Fig. 2B.

The field effects on the magnetization M' at different temperatures manifest a typical bulk superconductor as exemplified in the insets of Figs. 2A and 2B for 2 K. It shows that at 2 K, magnetic flux starts to penetrate into the sample at a field as low as 0.1 Oe. As a result, it is difficult to determine accurately the superconducting volume fraction of the 3UC FeSe-films, because of the large uncertainty in estimating the effects of demagnetization and flux penetration in such ultra-thin film samples (S2). The random variation of the values of both $\chi_{dc}$ and $\chi_{ac}$ of the 1-4UC FeSe-films prevents us from determining whether superconductivity originates only from the first mono FeSe-layer as previously reported, due to the uncertainty in sample growth and stability.

Our observations of flux penetration below ~ 0.1 Oe at 2 K in contrast to the large $H_{c2}$ and superconductivity up to ~ 80 K in contrast to the energy gap opening at 60±5 K previously reported suggest the possible existence of mesoscopic structures in the samples below and/or above their $T_c$. The results may also help explain the difficulty to produce FeSe films with high $T_c$. The frequency dependence of the diamagnetic susceptibility $\chi'_{ac}(\omega)$ should shed light on the above (S3). In the frequency range of $0.5 < \omega < 1.5$ kHz of our experiment, we found four temperature regions where $\chi'_{ac}$ decreases with different predominant ω-dependences as exhibited in Fig. 4 A-D, while $\chi''_{ac}$ increases linearly with ω. For T < 20 K, $\chi'_{ac} \propto \ln\omega$; for 20 K < T < 60 K, $\chi'_{ac} \propto \omega$; for 60 K < T < 100 K, $\chi'_{ac} \propto \ln\omega$; and for T > 100 K, $\chi'_{ac} \propto \omega^2$. From these ω-dependences of $\chi'_{ac}$, we conclude that below 20 K, the samples consist of weak-links, consistent with the low penetration field even at 2 K; between 20 and 60 K, the sample contains unconnected superconducting patches of dimension ≤ 1 μm, in agreement with the absence of the Meissner state; between 60 and 80 K, there appear unknown randomly distributed excitations of superconducting and magnetic nature; and above 100 K, no superconductivity exists. The analysis of the phase angle $\tan^{-1}(M''/M')$ appears in agreement with the suggestion above (S4). However, it should be noted that the cause of the heterogeneous nature of the FeSe thin films revealed by the present study remains unknown, in spite of its significant potential implications on the origin of high $T_c$ and nano-electronics.

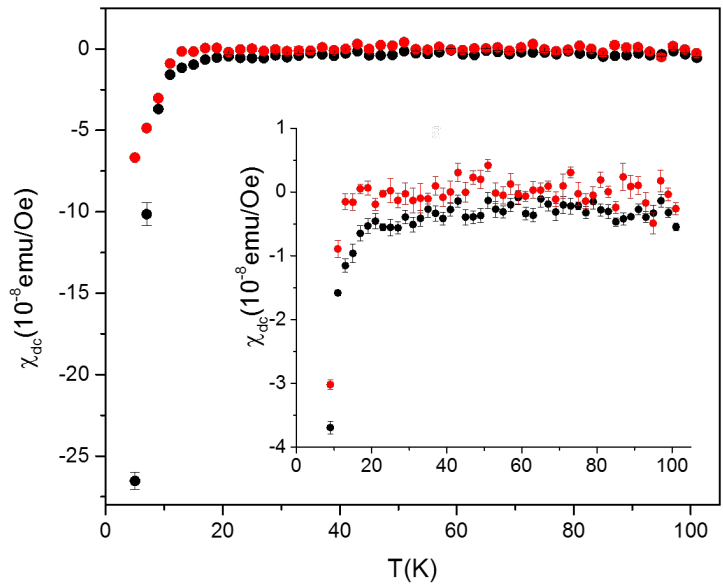

**Fig. 1.** The temperature dependence of ZFC-$\chi_{dc}$ and FC-$\chi_{dc}$ for the 3UC FeSe film in a dc field of 10 Oe. The error bars represent the data spreads over twenty independent measurements. The insert shows an expanded scale.

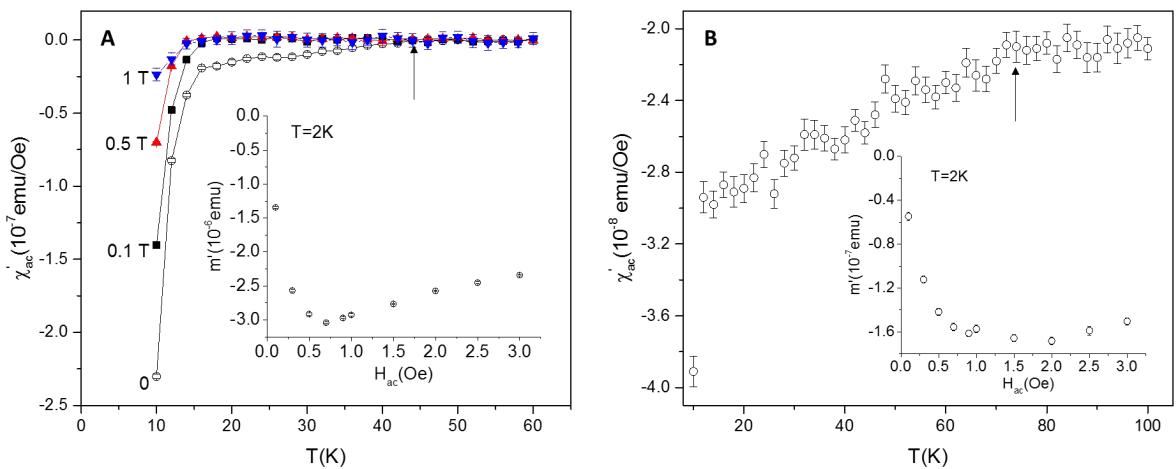

**Fig. 2.** The temperature dependence of $\chi'_{ac}$ at an ac field of 3Oe at 1.5 kHz for the 3UC-sample (A) and the 4UC-sample (B), with arrows showing the temperatures where diamagnetic shifts start. The inserts represent M'-H for $H_{ac}$ = 0-3Oe at 2K. The error bars represent the data spread over fifty independent measurements.

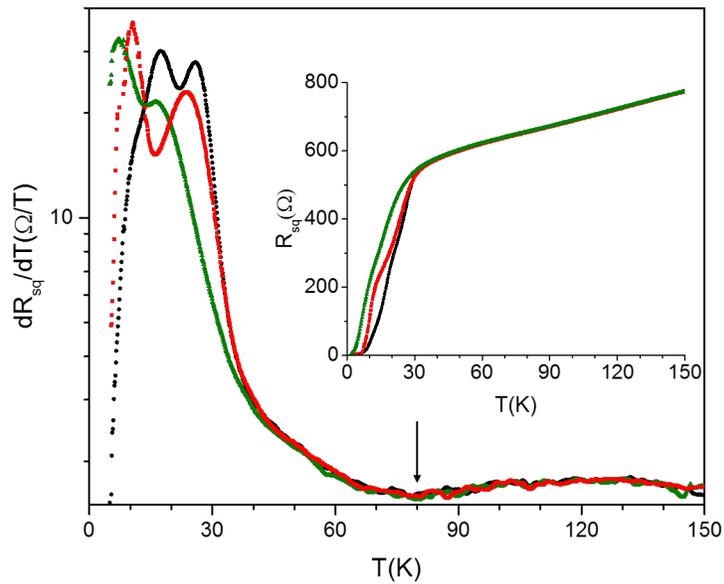

**Fig. 3.** Differentiation of resistance dR/dT of the 4UC FeSe film under different magnetic fields: ● 0T, ■ 1T and ▲ 7T, with the arrow showing where dR/dH starts to increase. Insert: resistance of the 4UC FeSe film from 2-150K under different magnetic fields.

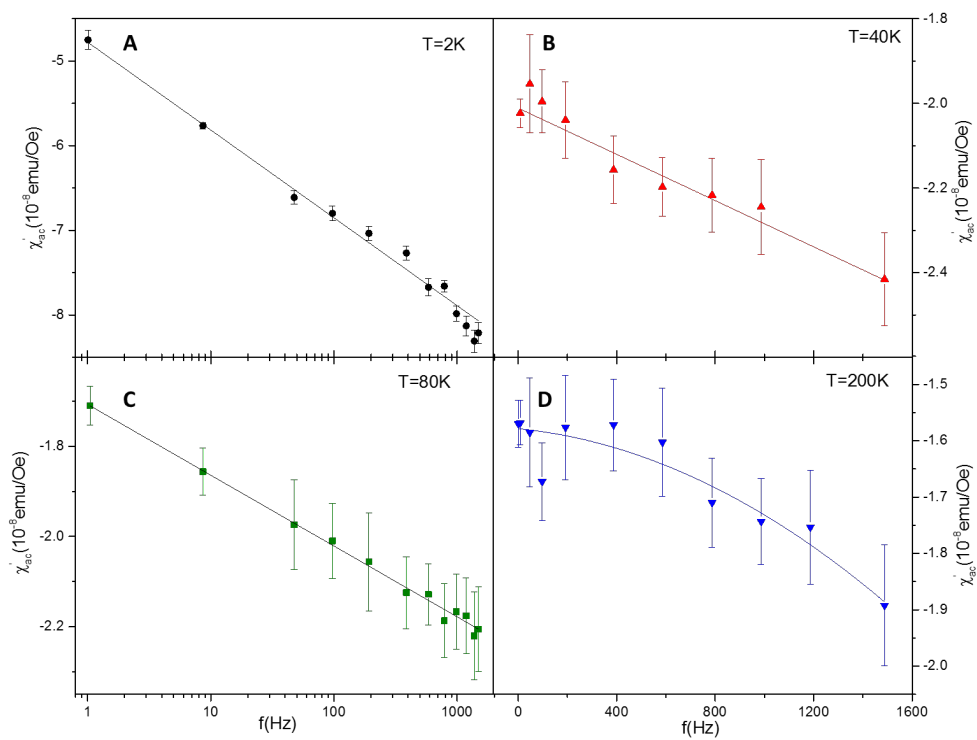

**Fig.4.** Representative frequency dependency of $\chi'_{ac}$ from 0.5 to 1500 Hz of ac susceptibility for the 4UC FeSe film at different temperature regions: e.g. at A) 2K, B) 40K, C) 80K, and D) 200K. The error bars represent the data spreads over fifty independent measurements for each data point. The solid lines represents the logarithmic (black), linear (red), and parabolic (blue) fitting of the data.

Supplementary Materials for:

The Meissner and Mesoscopic Superconducting States in 1-4 Unit-Cell FeSe-Films up to 80 K

L. Z. Deng[1], B. Lv[1], Z. Wu[1], Y. Y. Xue[1], W. H. Zhang[2], F. H. Li[2], L. L. Wang[3], X. C. Ma[3], Q. K. Xue[2] and C. W. Chu[1,4]

[1] Texas Center for Superconductivity and Department of Physics, University of Houston, Texas

[2] Department of Physics, Tsinghua University, Beijing

[3] Institute of Physics, Chinese Academy of Sciences, Beijing

[4] Lawrence Berkeley National Laboratory, Berkeley, California

1. **Magnetization data acquisition principle and procedures**

The mass of the ultra-thin nUC FeSe film samples are small ~ nx26ng [density=4.72g/cc, dimension=0.2cm x 0.5cm, 1UC thickness=5.525 x $10^{-8}$ cm]. The small $\chi_{dc}$ and $\chi_{ac}$ values presented have been subtracted with the background (Si/10UC FeTe/STO, ~$10^{-8}$ emu/Oe), which is T-independent and shows no hysteresis. To further exclude possible data fluctuations, we have performed 20 independent measurements for the $\chi_{dc}$ and 50 independent measurements for the $\chi_{ac}$ for each single data point from 2K to 100K. Chauvenet's criterion is used for rejecting bad data points. The error bars presented in the figures clearly show a data fluctuation 1-2 orders smaller than the magnitude of the signal obtained. Therefore, the data truly represent the signal from the FeSe films with ultra-small masses (n x 26 nanograms).

2. **Superconducting volume fraction estimations for the FeSe thin films**

The calculation of the volume fraction for superconducting thin films requires the exact information of the film's large demagnetization factor and the complex field penetration behavior, and both are difficult to estimate. Taking the 1UC FeSe with film area of 10 mm$^2$ as an example (which gives an equivalent diameter of 3.26 mm for the disk with d/l > 20 for the estimation), the demagnetization enhancement factor is D= 0.64d/l = 0.64x3.26/(5.525x10$^{-7}$) = 3.8x10$^6$ [1]. If there were no flux penetration, the superconducting volume fraction at 2K with applied field 3 Oe is estimated to be $f = 4\pi M = 4\pi \dfrac{M}{H \cdot V \cdot D}$ ~2%. Degrading sample showed high nonlinear MH below 3 Oe for the 3UC films shown in Fig. 2A. However, this estimation can be very rough since we cannot probe the exact interface volume (which can be much smaller than the estimation above) responsible for the superconductivity. The superconducting volume fraction could be up to 100% shielding fraction if there were a few possible weak-links, which will significantly suppress the demagnetizing enhancement. In addition, as we mentioned in the main text, the apparent flux penetration field of these FeSe thin films is very low (expected to be far below 0.001 Oe). This will make the volume fraction estimation even more complicated.

## 3. The relationship between m' and frequency ω in the FeSe thin films

The ω-dependence of $\chi_{ac} = \chi'_{ac} + i\chi''_{ac}$ for a normal metal has been given by the Landau theory [2] as $\chi' \propto \omega^2$ and $\chi'' \propto \omega$. For homogeneous superconductors, m' should be diamagnetic ω-independent [3-4]. The corresponding phase angle at the low-ω limit will be 90° and 180°, respectively.

However, the observed $\chi_{ac}$ of the FeSe ultra-thin films are qualitatively different. In the FeSe thin film, four regions with different ω dependences (and the phase angles θ at the low-ω limit) are clearly observed: for T < 20 K, $\chi'_{ac} \propto \ln\omega$; for 20 K < T < 60 K, $\chi'_{ac} \propto \omega$ and θ ≈ 115°~140°; for 60 K < T < 100 K, $\chi'_{ac} \propto \ln\omega$ and θ ≤ 90°; and for T > 100 K, $\chi'_{ac} \propto \omega^2$ with θ ≈ 90°. The $\chi'_{ac} \propto \omega^2$ indicates the normal metal properties of the FeSe film above 100K. The observations of $\chi'_{ac}(\omega)$ below 100K could be understood as following:

1) The clear logarithmic ω dependency below 20 K resembles a superconductive glass state. The existence of such glass state at $H_{dc}$ < 1 mOe further suggests a possible origin of sample inhomogeneity instead of the well-known vortex glass.

2) The superconductivity for 20 K < T < 60 K can be clearly observed from ac susceptibility measurements while hardly detectable from the dc susceptibility measurements (lack of persistent supercurrent). This suggested that the samples are a mesoscopic mixture of the normal metal and superconductor far below the percolation threshold with well separated superconducting patches. Therefore, no noticeable Josephson tunneling currents are allowed. The possible superconductive domains should also be far smaller than the respective London penetration depth. Under the *ac* driven field, however, the eddy currents across adjacent patches may lead to the circulating currents across both the patches and the normal-metal sections. Estimations further demonstrate, consequentially, that an apparent $\theta$ between 90° and 180° as well as a linear dispersion $\chi \propto \omega$ will be possible under such configuration.

3) A m'$\propto$ ln$\omega$ relationship is observed again in the range of 60 K < T < 100 K, similar to the observation below 20K. This implies the existence of possible glass-like spin excitations, whose moment increases with cooling but disappears again below 40-60 K. However, such dominating component may not exclude possible minor superconducting contribution, which is linear on $\omega$ as discussed above. Actually, both the dR/dT onset (Fig. 3) and the diamagnetic drop (Fig. 2B) suggest the existence of such minor phase up to 80 K. A decisive verification is yet to be made.

4. **The relationship of phase angles with temperature for normal metal and superconductors.**

As mentioned above, for normal metals, the induced current is driven by the induced electric field E $\propto -\frac{\partial B_{ac}}{\partial t} \approx i\omega H_{ac,ext}$. At low enough frequencies, the phase angle $\theta$ is expected to be 90°.

For superconductors, the persistent supercurrent is driven by the vector potential A, which is in-phase with $B_{ac}$ based on the correlation of B = $\nabla$XA. Therefore, a phase angle $\theta = 180°$ is expected.

For patch-like superconductors, different parts along a current loop may be made of both superconductive sections and non-superconductive ones that will have to accommodate the different phase-angles demented by different sections. Model calculations further demonstrate that a self-consistent solution exists with the circulating current proportional to ω and the phase-angle between 90° and 180°.

The above has been demonstrated in Fig. S1 for simple normal metal Ag, a simple superconductor Pb, and a high temperature superconductor YBCO film with spin excitations up to 200K, in comparison with observations in the 4UC FeSe-sample.

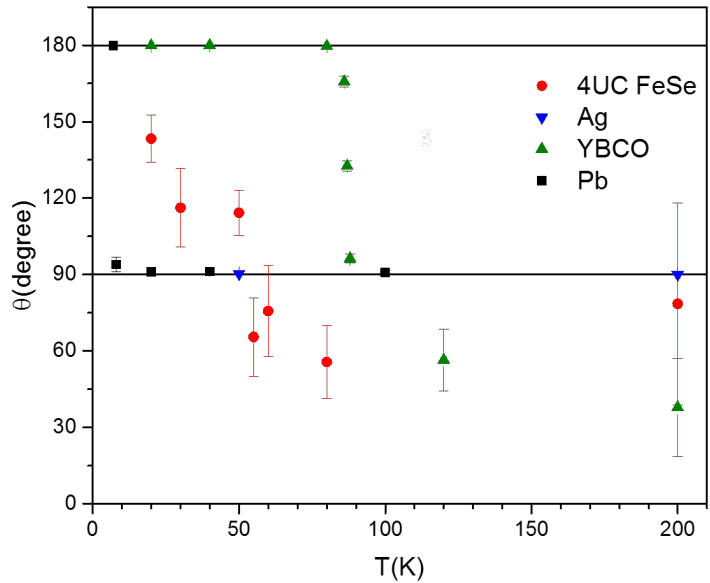

**Fig. S1.** Temperature dependency of phase angle for normal metal Ag foil (▼), superconductor Pb (■), YBCO thin film (▲), and 4UC FeSe thin film (●).

References

[1] R.M. Bozorth and D. M. Chapin, J. Appl. Phys. 13,320 (1942).

[2] *Electrodynamics of continuous media*, Landau and Lifshitz, pp. 199-208.